\documentclass[10pts]{kluwer}
\usepackage{epsfig}
\usepackage{epsf}
\def\G1915{GRS $1915$+$105$}
\def\X1550{XTE J$1550$-$564$}
\def\J1655{GRO J$1655$-$40$}
\def\etal{{\em et al. }}
\begin{document}
\begin{article}
\begin{opening}
\title{The Accretion Ejection Instability : Observational Tests}
\author{J. Rodriguez \email{jrodriguez@cea.fr}}
\author{P. Varni\`ere}
\author{M. Tagger}
\author{Ph. Durouchoux}
\institute{Service d'Astrophysique (CNRS URA 2052)
,CEA/Saclay 91191 Gif sur Yvette}
\runningauthor{Rodriguez \etal}
\runningtitle{AEI: Observational Tests}
\end{opening}

\section{First Steps : \J1655}
QPOs are commonly considered to originate in the disks of X-ray
binaries.  One thus expects, and most often observes, that the QPO
frequency decreases as the disk inner radius increases.  However Sobczak
\etal (1999) pointed out an opposite behavior in the case of \J1655.
This might prove usefull in testing the theory of the
Accretion-Ejection Instability (see contributions by Caunt, Tagger and
Varni\`ere, these proceedings). At
the same time, Merloni, Fabian, and Ross (1999) showed that, when
fitting the sources spectra with the standard model (multicolor
blackbody+ power law), severe distorsions could be due to the vertical
structure of the disk.  In that view we re-did the spectral analysis of
the observations in which Sobczak \etal found QPO's.  \\
Using the
criterion of Merloni \etal, we find that all the data points from \J1655
where the fits give an anomalously low inner radius must be excluded.
The remaining points confirm the inverse correlation.
\section{Another Good Candidate : \G1915}
Because of its high and rapid variability, and following the results
found by Markwardt et al., 1999, and Muno et al., 1999, we then focussed
on \G1915.  In particular we studied very precisely two 30 min
cycles present in the observation of September $9^{th}$ 1997, in which
the QPO frequency had been seen to vary (Markwardt \etal, 1999, Swank
\etal, 1997), correlated to the flux, the color temperature and the
color radius itself.

We extracted, during two $\sim 30min$ dips, every 20s a 16s spectrum and
a lightcurve, extracting both the QPO frequency and the spectral
parameters (Rodriguez \etal, 2000).

We have also analyzed recent observations of \G1915, pointed as a target
of opportunity, on April $17^{th}, 22^{th}, 23^{th}$ by RXTE. Although
only the first date showed high variability, in the three observations a
strong QPO, together with its first harmonic are highly correlated to
the flux (work in progress).

\section{Results and Interpretation}

\begin{itemize}
\item Applying more stringent criteria leads us to exclude some of the
data points observed by Sobczak \etal in  \J1655, but the points we
retain do show the inverse correlation between the QPO frequency and the
disk inner radius.

\item Though highly variable \G1915 does not present this inverse
correlation, during the 30 mn cycles.

\item These observations are compatible with the theory of the
Accretion-Ejection Instability (see Varni\`ere., these proceedings), if
the disk of \J1655 is always near the Last Stable Orbit when QPO are
observed, and the disk of  \G1915 never approaches it during its 30 mn
cycles.

\item The criterion of Merloni et al.  leads us to exclude the data
points of \J1655 where the spectral fits return a very low color
radius($\sim$ very few kms).  But we find that these points are not
compatible with the predictions of Merloni, as they systematically show
a higher temperature when the color radius decreases.  On the other
hand, data points at the onset of the low state, in \G1915, are
compatible with the predictions of Merloni \etal.

\item We note that in \J1655, as well as during the hard-steady state
of \G1915, the very low color radius actually means a small size of
the emitting region.  This and the higher temperature would be
compatible with the presence in the disk of a spiral shock or hot point,
where accretion energy would be dissipated.

\end{itemize}

\end{article}

\end{document}